\documentclass{PoS}

\usepackage{graphicx}
\usepackage{subfigure}

\usepackage{amsmath}
\usepackage{comment}

\title{Methods for evaluating physical processes in strong external fields at $e^+e^-$ colliders: Furry picture and quasi-classical approach}

\ShortTitle{Methods for evaluating physical processes in strong external fields at $e^+e^-$ colliders}

\author{\speaker{Stefano Porto}\thanks{S.~P. gratefully acknowledges support of the DFG throught the grant SFB 676 ''Particles, Strings, and the Early Universe''.}\\
        II. Institut f\"{u}r Theoretische Physik, University of Hamburg, Luruper Chaussee 149, D-22761 Hamburg, Germany
\\
        E-mail: \email{stefano.porto@desy.de}
}

\author{Anthony Hartin\\
        DESY, Deutsches Elektronen-Synchrotron, Notkestr. 85, D-22607 Hamburg, Germany
}

\author{Gudrid Moortgat-Pick\\
        II. Institut f\"{u}r Theoretische Physik, University of Hamburg, Luruper Chaussee 149, D-22761 Hamburg, Germany; \\DESY, Deutsches Elektronen-Synchrotron, Notkestr. 85, D-22607 Hamburg, Germany
}

\abstract{
Future linear colliders designs, ILC and CLIC, are expected to be
  powerful machines for the discovery of 
  Physics Beyond the Standard Model and subsequent precision studies. 
However, due to the intense beams (high
  luminosity, high energy), strong electromagnetic fields occur in the
  beam-beam interaction region. In
  the context of precision high energy physics, 
the presence of such strong fields
  may yield sensitive corrections to the observed electron-positron
  processes. The Furry picture of quantum states gives a conceptually
  simple tool to treat physics processes in an external field. A
  generalization of the quasi-classical operator method (QOM) as an approximation is
  considered too.
}

\FullConference{Proceedings of the Corfu Summer Institute 2012 "School and Workshops on Elementary Particle Physics and Gravity"\\
		September 8-27, 2012\\
		Corfu, Greece}

\begin{document}


\section{Introduction}

After the first successful data runs of the LHC a completely new energy
region at the Terascale energy frontier has been touched and already
one new boson, a Higgs boson, has been discovered.  Thanks to the
excellent luminosity and detector performance at the LHC, already 
a rather precise and consistent mass measurement of the new particle
has been achieved,
\begin{equation}
m_{\rm Higgs}\sim 126 \,\,\mbox{\rm GeV}\,\,\,\cite{ATLAS-CONF-2013-014,CMS-PAS-HIG-13-002}.
\end{equation}
Such a mass for a Higgs boson is perfectly in agreement with mass
predictions from high precision analyses in the electroweak
sector. However, in order to clearly manifest whether the discovered
particle is the Standard Model (SM) Higgs boson or whether the candidate
points to an extended model as, for instance, Supersymmetry, a precise
measurement of all couplings, the total width and branching ratios is
necessary. Therefore corresponding measurements at an $e^+e^-$ collider
are highly desirable.

Designs for a planned future $e^+e^-$ linear collider (LC) are set up to
reach high $\sqrt{s}$ up to 1-1.5 TeV for the ILC and about 3 TeV for
CLIC, with a very high
luminosity in the range of $10^{-34}$-$10^{-35}$ cm$^{-2}$s$^{-1}$
\cite{ILC-tdr,clic-cdr}.

The key arguments for a linear lepton collider are, for instance,
 the clean environment
 and the availability of polarized beams in the initial state, as well
 as rather low background rates 
compared with corresponding rates at
a hadron collider.
These facts permit in
principle a full reconstruction of the observed 
processes and enable unprecedented high precision measurements.
Therefore  the LC is
expected to perform high sensitivity for precision physics in the Higgs boson,
top quark and electroweak gauge bosons sectors and in both direct, as well as
indirect searches for new physics Beyond the Standard Model (BSM).
Thus, the planned linear collider physics potential can strikingly
contribute and even extend the high energy physics range of the LHC. 

In order to reach the high luminosity required by the comprehensive
physics program, the LC applies strongly squeezed and intense bunches of
electrons and positrons at the Interaction Point (IP). At the ILC, for instance, about
a factor 1000 more intense beams, i.e. more $e^-$/$e^+$ per pulse, will be achieved than at the SLC. In order to
optimize the outcome
of such high precision physics at a LC, one needs to know, however, in detail
all processes occurring at the IP.

Each charged bunch generates an intense collective electromagnetic
field, that is felt by the interacting particles from the oncoming
beam. Due to its intensity, this bunch field can be considered as one
external electromagnetic field as a whole rather than to be composed by
single photons interacting with the colliding leptons.


We study the impact of such external fields on the actual physics
processes in electroweak and BSM physics. To achieve this goal, a full
comprehension of QED effects in an external field (non linear QED) is required.

Processes in external electromagnetic fields have caused
interest since the beginning of quantum electrodynamics, with the
paradox of electron quantum tunneling in an arbitrary high potential
barrier, observed by Klein in 1929 \cite{Klein}. Sauter (1931) \cite{Sauter}
showed that this effect depends exponentially on the intensity of the
field in the barrier. Schwinger (1951) \cite{Schwinger} interpreted this
paradox via the concept of a critical field: the Schwinger critical
field ($E_{cr}=m_e^2/e\simeq$1.3$\cdot 10^{18}$ V/m,
$B_{cr}=m_e^2/e\simeq4.41\cdot 10^{9}$ T, using natural units)
corresponds to the intensity of an electromagnetic field at which the
vacuum spontaneously creates electron-positron pairs. The vacuum that is
due to the external field full of \textit{virtual} particles becomes
unstable and transforms into a more stable charged vacuum via producing
\textit{real} particles \cite{FradkinBook}.

The critical field can be reached in nature on the surface of pulsars
and magnetars \cite{Kuznetsov:2004tb} and close to superheavy nuclei
\cite{Greiner}. Such conditions are very difficult to
reproduce in a laboratory. In recent years, new experiments have been
developed and designed where the Schwinger critical field condition could be
achieved in the boosted rest frame of a test electron. In the context of
low energy physics, this is for example the case for newly designed
intense lasers like XFEL at DESY \cite{XFEL}. Regarding high
energy physics (HEP), the first experiment to study the strong field regime of
nonlinear QED was the E-144 experiment at SLAC \cite{Bamber}, in which 46.6 GeV electrons were shot through an intense laser. A second
example of sources for non linear QED effects
in HEP experiments are, as explained above, the interaction points (IPs) at future
linear collider designs, ILC \cite{ILC-tdr} and CLIC \cite{clic-cdr}. The Schwinger
condition may be fulfilled, indeed, in the rest frame of the ultrarelativistic
particles that scatter at the IP.

Quantum processes in intense electromagnetic fields have been
studied in different research areas like intense plasma and laser
physics \cite{Heinzl}
(for a detailed review see also \cite{DiPiazza12}), using the
so called Furry picture (FP) \cite{Furry}. According to the FP, the
external field is treated as a classical object that modifies the
equations of motions. However, an analogous appropriate study is still
lacking for the case of linear colliders, and this is the aim of our
studies.

In these proceedings, we will address the question whether electroweak
physics processes may be affected by the presence of the very intense
external electromagnetic fields at the IPs at colliders 
and whether these effects can be
easily incorporated in the calculations.

In section \ref{Section_Linacs} we will introduce the parameters that
need to be considered in general in the presence of an external
electromagnetic field, describing also the generated fields at
the interaction points of a future linear collider; in section
\ref{Section_FurryPicture} we introduce the Furry picture formalism and
explain how to apply it to the discussed processes; in
section \ref{Section_QOM} we sketch the Ba\u{\i}er-Katkov quasi-classical
operator method as a possible alternative in the case of
ultrarelativistic particles; we conclude in section \ref{Conclusions}.


\section{Intense fields at linear colliders}\label{Section_Linacs}

In order to perform the high precision physics program planned for
future linear colliders, very high luminosity $\mathit{L}$ is needed.
Therefore extremely squeezed $e^+$ and $e^-$ bunches are required, 
$$\mathit{L}\propto\frac{N_{e^+}N_{e^-}}{\sigma_x\sigma_y},$$ where
$N_{e^{\pm}}$ is the number of $e^{\pm}$ per bunch and $\sigma_x$,
$\sigma_y$ are the transversal dimensions of the bunch propagating along
$z$.

Each of the  dense bunches ($N_{e^{\pm}}\sim10^{10}$ and $\sigma_x$,
$\sigma_y\sim 1$-$10^3$ nm) can be regarded as an electromagnetic current
generating a collective strong electromagnetic field at the IP.  

Correspondingly, the colliding particles at the IP will see a
superposition of the collective fields originating from the two beams.
Due to the boosted particle frame, these fields can be approximated by
two almost anticollinear constant crossed fields; each particle will
mainly see the constant crossed field from the oncoming bunch
\cite{YokoyaChen}.
A constant crossed field is the limit
for infinite period of a plane wave field with momentum
$k=\frac{c}{\omega}$, it has a trivial spatial dependence
$A^{\mu}(k\cdot x)=a^{\mu}\,k\cdot x$ and its electric ($\mathbf{E}$)
and magnetic ($\mathbf{B}$) components are orthogonal and equal in
magnitude:
\begin{equation}\label{ConstCrossF}
 \mathbf{E} \bot\mathbf{B},\,\,\,\,\,\,\,\,|\mathbf{E}|=|\mathbf{B}|.
\end{equation}

Due to momentum transfer, the presence of external electromagnetic field
permits processes that are kinematically not allowed, as, for instance,
beamstrahlung (i.e. bremsstrahlung in the electromagnetic field of a
relativistic particle bunch) and coherent pair production.  These
external fields can also affect the rate of the 
allowed processes like
incoherent pair production.

At previous accelerators LEP and SLC these effects have been
considered and estimated with some approximations. Beamstrahlung and
coherent pair production have been treated via the Ba\u{\i}er-Katkov
quasi-classical operator method (QOM) \cite{Baier1968} while incoherent
pair production was described via the equivalent photon approximation (EPA)
\cite{EPA}.

However, effects from strong external fields do not only affect the previously
mentioned ``background'' processes at the IP, but they may also have direct
consequences on the electroweak SM or BSM processes that are in the focus
of future colliders. To our knowledge,
this problem has never been addressed at colliders. Only some topics have
been discussed in the context of laser physics
(\cite{Muller, DiPiazza12} and references therein), astroparticle physics~\cite{Kuznetsov:2004tb} and decays in extremely intense fields \cite{Kurilin:1999qc, Nikishov64B}.

Therefore a correct knowledge of the QED effects within a strong external field
environment is required
in order to optimize the
physics program with the expected precision.

For the description of the external electromagnetic field at the IP, it
is useful to introduce four Lorentz and gauge invariants that one
can compose with  the external field strength tensor $F_{\mu\nu}$ and
the momenta of the propagating particle $p_{\mu}$. The considered propagating particle can also be a
photon. The probabilities of the processes with a \textit{single} initial
particle in a general external field (ex. beamstrahlung, coherent pair
production) depend on the following quantities \cite{RitusNauka}:
\begin{align}
x&=\frac{e\sqrt{(A_{\mu})^2}}{m}=\frac{eE}{m\omega}\\
 \chi&=\frac{e}{m^3}\sqrt{(F_{\mu\nu}p^{\mu})^2}\\\label{F_parameter}
F&=\frac{1}{4}F_{\mu\nu}F^{\mu\nu}=\mathbf{B}^2-\mathbf{E}^2\\ 
 |G|&=\lvert\frac{1}{4}F_{\mu\nu}\tilde{F}^{\mu\nu}\rvert=|-\mathbf{E}\cdot \mathbf{B}|\label{G_parameter}
\end{align}


where $m$ and $e$ denote the mass and the charge of the propagating
particle (for an initial photon one uses the mass and charge of the
electron) and
$\tilde{F}^{\mu\nu}=\frac{1}{2}\epsilon^{\mu\nu\rho\sigma}F_{\rho\sigma}$.

 The parameter $x$ represents the work done by the external field in a Compton length
$\lambda_C=eF\hbar/mc$ in units of the energy $\hbar\omega$ of the
quanta of the external field. Therefore, for $x\ll1$, a low number of photons from the external field are expected to interfere with the interacting particles, while for $x\gtrsim1$, multiphoton processes are favoured with a nonlinear dependence on the external
field. Therefore $x$ is called the classical 
(since it does not depend on $\hbar$) nonlinearity parameter.

The $\chi$ parameter, often also called $\Upsilon$ in the literature, 
represents in units of $mc^2$ the work performed by the field
over the Compton length in the particle rest system. It parametrizes the
magnitude of the quantum nonlinear effects and is called the quantum
nonlinearity parameter \cite{DiPiazza12}. In particular, in the case of
ultrarelativistic particles, it describes the intensity of the external
field in the particle frame in units of the Schwinger critical
field:$$\chi=\frac{\gamma_L E}{E_{cr}}=\frac{\gamma_L B}{B_{cr}}$$ where
$\gamma_L$ denotes the Lorentz factor and $E$, $B$ the electric and magnetic components in
the laboratory frame. Highly energetic initial particles can see a
critical regime also in less intense fields in the laboratory frame, as
long as $\chi\sim$ $O(1)$. This value would correspond to an external
field of the order of the Schwinger critical field, at which the vacuum is
polarized.
The parameters $F$ and $|G|$ instead describe respectively the relative magnitude and orientation between $\mathbf{E}$ and $\mathbf{B}$.

The probability $W$ of a process with one initial particle in a constant
background field depends in general only on $\chi, F, |G|$ since
correspondingly $x\gg1$ \cite{RitusNauka}. According to
\eqref{ConstCrossF}, \eqref{F_parameter} and \eqref{G_parameter}, one
has  $F, |G|=0$ for a constant crossed field.  In the case of an
ultrarelativistic particle ($p^0\gg m_e$) in a relatively weak field
compared to $E_{cr}$, one has $|F|,|G|\ll \mbox{min}(1,\chi^2)$
\cite{RitusNauka}, confirming that a crossed field is a good
approximation for the field seen by the colliding particles at the IP of
LCs.  One can generalize the above picture also to processes with two
initial particles in an external field.  As a consequence, the
probabilities of processes at the IPs simplify $W(\chi, F, |G|)\simeq
W(\chi, 0, 0)$, depending effectively only on the intensity of the
external field. Moreover, constant crossed fields allow simpler
analytical calculations and integrations and have been object of recent
studies \cite{King}.

The $\chi$ parameter varies during the collision since the bunches
distort under the pinch effect and the disruption effects. According to
\cite{YokoyaChen}, the average value for $\chi$ in a
gaussian bunch is given by:
\begin{equation}\label{YokoyaApprox}
  \chi_{average}\approx \frac{5}{6}\frac{Nr^2_e\gamma_L}{\alpha_{em} \sigma_z(\sigma_x+\sigma_y)}
\end{equation}
where $N$ is the number of leptons per oncoming bunch, $\alpha_{em}$ the
fine structure constant, $r_e$ the Compton radius, $\sigma_x,\sigma_y$
the transversal dimensions of the bunches, $\sigma_z$ the longitudinal
dimension.  Using the IP beam-beam simulation program \texttt{CAIN}
\cite{CAIN} one obtains the values presented
in Tab. \ref{Parameters} for ILC and
CLIC, while for LEP II and SLC we used the approximations~\eqref{YokoyaApprox}.

\begin{table}\begin{center}
\begin{tabular}{|c||c|c|c|c|}
\hline  
Machine & LEP II & SLC & ILC-1TeV& CLIC-3TeV\\ \hline\hline
 Energy (GeV) & 94.5 & 46.6 & 500 & 1500\\ \hline
 $N$ (10$^{10}$) & 334 & 4 & 2 & 0.37 \\ \hline
 $\sigma_x,\sigma_y$ ($\mu$m) & 190, 3 & 2.1, 0.9 & 0.49, 0.002 & 0.045, 0.001\\ \hline
 $\sigma_z$ (mm) & 20 & 1.1 & 0.15& 0.044\\ \hline\
 $\chi_{average}$  & 0.00015 & 0.001 & 0.27 & 3.34\\ \hline
\end{tabular}\end{center}
\caption{Lepton colliders parameters. $N$ is the number of leptons per
bunch, $\sigma_x,\sigma_y$ are the transversal dimensions of the bunches,
$\sigma_z$ presents the longitudinal dimension. $E$ is the energy of the
particles in the bunches. The parameters for ILC-1TeV are taken from a 2011 dataset.}  \label{Parameters}
\end{table}

It is clear that the electromagnetic fields at the IPs of future linear
colliders would be much more intense than in the previous lepton
accelerators\footnote{At the LHC, the less
dense bunches do not allow to reach the really high critical field corresponding to proton pair production ($10^{24}$
V/m).}, with values of $\chi$ up to order
$O(1)$. Such a value for $\chi$ corresponds to the polarization of the
vacuum. This unstable vacuum requires incorporation within the
calculations in a non-trivial way. A study of the effectiveness of the
previously mentioned quasi-classical and EPA approximations for processes
in this physical case should be undertaken.

The expected $e^+e^-$ pair productions at linear colliders are listed in Tab. \ref{PairsProducted}.

\begin{table}
 \begin{center}
\begin{tabular}{l|c|c|}
\cline{2-3}
&ILC-1TeV& CLIC-3TeV\\ \hline
\multicolumn{1}{|l|}{$\#$ of  coherent pairs}&$\sim0$&$6.8\cdot10^{8}$\\ \hline
\multicolumn{1}{|l|}{$\#$ of  incoherent pairs}&$3.9\cdot10^{5}$&$3.8\cdot10^{5}$\\ \hline
\end{tabular}\end{center}
\caption{Pair production processes at the IP regions at ILC and CLIC.}
\label{PairsProducted}
\end{table}


\section{The Furry picture and its application}\label{Section_FurryPicture}
 
The intense external field experienced by each particle at the IP is
characterized by a high photon density and a corresponding wave function
overlap, so that it can be seen as a \textit{classical} external background
field. For physics in intense fields, the effects of such a classical
external field are taken into account \textit{exactly} through the so
called \textit{Furry picture} or representation (FP) of quantum states
\cite{Furry}.  The main idea of FP is to find the exact solutions of the
equation of motion in the external field, taking into account the latter
non perturbatively. Then, one applies the solutions in the
Feynman-Schwinger-Tomonaga S-matrix theory \cite{LandauQED} to calculate
the probabilities of the physical processes. We propose to use FP also
to calculate the probabilities of processes at the IP of linear
colliders. In the following, we briefly review the FP 
technique~\cite{MoortgatPick:2009zz}.

In the usual Interaction (or Dirac) picture the time dependence is
shared between the state vectors and the observables. In particular, the
Hamiltonian is given by 
\begin{equation}
\mathcal{H}=\mathcal{H}_0+\mathcal{V}
\end{equation}
where $\mathcal{H}_0$ is the time-independent unperturbed Hamiltonian,
describing the time evolution of observables, and $\mathcal{V}$ is the
interaction Hamiltonian, that regulates the time dependence of the
states. In QED, $\mathcal{V}$ represents the gauge interactions between
fermions and photons. The eigenstates of $\mathcal{H}_0$ are assumed to
be the free states of the particle in the vacuum.

In the FP the Hamiltonian is given by 
\begin{equation}
             \mathcal{H}=\mathcal{H}_0+\mathcal{H}_{ext}+\mathcal{V}=\mathcal{H}_B+\mathcal{V}
  \end{equation}
where $\mathcal{H}_{ext}$ represents the interaction of the fermions
with the external \textit{classical} field. In the FP the considered
state vectors are the bound states of the fermions in the external
field, eigenstates of $\mathcal{H}_B$. The FP bound eigenstates are
related by a canonical transformation to the free particle states of the
Interaction picture \cite{Furry}, they obey to different commutation
relations but, in the limit of null external field, the usual
commutation relations are recovered.  The corresponding QED Lagrangian
can be written as:\begin{equation}\label{FurryLagrangian}
\mathcal{L}=\bar{\psi}(i\displaystyle{\not}{\partial}-e\displaystyle{\not}{A}_{ext}-m)\psi-\frac{1}{4}FF-e\bar{\psi}\displaystyle{\not}{A}\psi
                                                   \end{equation}
where $A^{\mu}_{ext}$ is the classical external field and its interaction term is distinguished from the usual gauge interaction term. Note that there is no kinetic therm $\frac{1}{4}F_{ext}F_{ext}$ since $A^{\mu}_{ext}$ is a classical external background field, not a dynamical field.

From the Lagrangian \eqref{FurryLagrangian} one derives the modified Dirac equation:
\begin{equation}\label{VolkovEq}
 (i\displaystyle{\not}{\partial}-e\displaystyle{\not}{A}_{ext}-m)\psi=0
\end{equation}
the solution of which was found by Volkov in the '30s \cite{Volkov}:

\begin{equation}\label{VolkovSolution}
  \Psi^{V}_p(k\cdot x)=\frac{1}{\sqrt{(2\pi)^32\epsilon_p}}\,E_p(k\cdot x)\,\,u(p)
\end{equation}
with
\begin{equation}
   E_p(k\cdot x)\equiv\left(1-\frac{e\displaystyle{\not}A_{ext}\displaystyle{\not}k}{2(k\cdot p)}\right)\mbox{exp}\left[-ip\cdot x-i\int^{(k\cdot x)}_0\left[\frac{e(A_{ext}(\phi)\cdot p)}{(k\cdot p)}-\frac{e^2A_{ext}(\phi)^2}{2(k\cdot p)}\right]d\phi\right],
\end{equation}
where $k$ is the momentum of the external field, $p$ and $\epsilon_p$
the canonical momentum and energy of the fermion while $u(p)$ is the
usual Dirac spinor solution.

The solution \eqref{VolkovSolution} is valid for a vast class of
classical external fields. However, the precise expression is known only
for few configurations like the plane wave electromagnetic field, the
crossed electromagnetic field and the Coulomb field. This solution takes
entirely into account the effects of the external electromagnetic field
on the fermion.  The $\Psi_V$ solutions constitute an orthogonal and
complete system \cite{Ritus1970}, for a review \cite{Boca}.  Solutions
of analogue equations of motion for charged scalars and vector bosons in
an external field have been found, see \cite{Kurilin:1999qc} for a
review.

$\Psi_V$ and the analogue scalar and vector solutions can be used in perturbation theory to build new Feynman rules and diagrams in order to describe processes in an external field.

The described solutions of the modified Dirac equation \eqref{VolkovEq}, of the analogue modified Klein-Gordon equation etc. can be used in perturbation theory to build new Feynman rules and diagrams in order to describe processes in an external field.

\begin{center}
 \begin{figure*}[h]
  \centerline{
    \mbox{\includegraphics[width=2.0in]{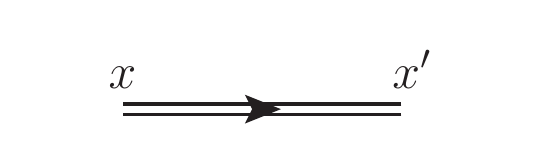}}
      }
  \caption{The FP fermion propagator. The double line represents the Volkov solution.}
  \label{FP_Prop}
  \end{figure*}
\end{center}

In particular, here we give the QED FP-Feynman rules at the tree-level, built out of solution \eqref{VolkovSolution}. The fermion two-point function in coordinate space is given by, see Fig. \eqref{FP_Prop}:
\begin{equation}
\label{FP_prop_fermion}
 G(x,x^{\prime})=\frac{1}{(2\pi)^4}\int_{-\infty}^{+\infty}d^4p\,\,E_p(k\cdot x)\frac{\displaystyle{\not}{p}+m}{p^2-m^2}\,\overline{E}_p(k\cdot x^{\prime})e^{ip\cdot(x^{\prime}-x)},
\end{equation}
where the usual fermion propagator is sandwiched between $E_p$ factors ($\overline{E}_p=E_p^{\dagger}\gamma^0$) coming from the Volkov solutions. It is interesting to note that there is a non trivial dependence on the coordinates $x$, $x^{\prime}$ within which the fermion propagates, instead of their difference $x^{\prime}-x$ \cite{LandauQED}.

\begin{center}
 \begin{figure*}[h]
  \centerline{
    \mbox{\includegraphics[width=5.9in]{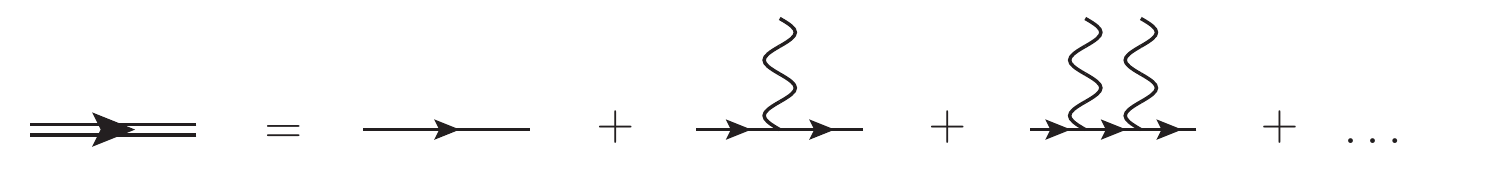}}
      }
  \caption{Interpretation of the electron propagator derived from the Volkov solution.}
  \label{FP_PropInter}
  \end{figure*}
\end{center}

Often in laser physics \cite{Heinzl}, 
the FP fermion line is interpreted considering a bare fermion
``dressed'' 
by an arbitrary number of photons emitted or absorbed from the 
 laser, Fig. \ref{FP_PropInter}.

\begin{center}
 \begin{figure*}[h]
  \centerline{
    \mbox{\includegraphics[width=2.0in]{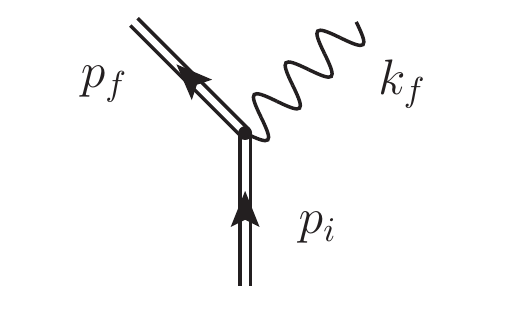}}
      }
  \caption{The FP QED vertex.}
  \label{QED_vert}
  \end{figure*}
\end{center}

The QED vertex in Fig. \eqref{QED_vert}, instead, is given by:
\begin{equation}\label{FP_vert_SUM}
 -ie\,\gamma^e_{\mu}=-ie\,(2\pi)^4\sum^{+\infty}_{r=-\infty}\, \overline{E}_{p_f}(r)\gamma_{\mu}E_{p_i}(r)\,\,\,\delta^4(p_f+k_f-p_i\;-\;r\,k).
\end{equation}
Each term of the sum is given by the usual Dirac matrix $\gamma^{\mu}$, sandwiched between the factors $E_p$, $\overline{E}_p$ and multiplied by a $\delta$-function with a momentum conservation law as argument. The latter contains a term $-rk$ that represents the momentum exchanged with the external field. 

In the case of a constant crossed field the sum in the vertex becomes an integral:
\begin{equation}\label{FP_vert_INT}
 -ie\,\gamma^e_{\mu}=-ie\,(2\pi)^4\int^{+\infty}_{-\infty}dr\,\,\, \overline{E}_{p_f}(r)\gamma_{\mu}E_{p_i}(r)\,\,\,\delta^4(p_f+k_f-p_i\;-\;r\,k).
\end{equation}

Initial and final (anti)fermions are described by the usual Dirac
spinors $u_p,\bar{u}_p, v_p, \bar{v}_p$ since the $E_p$ factors have
been grouped from the Volkov spinors in Feynman rules
\eqref{FP_vert_SUM} and \eqref{FP_vert_INT}.  The photon propagator at
tree level is unchanged, since the photon has null charge and it does
not interact directly to the external field.

\begin{figure}[htb]\centering
\subfigure[Beamstrahlung]{\label{Beamstrahlung1}\includegraphics[width=.35\textwidth]{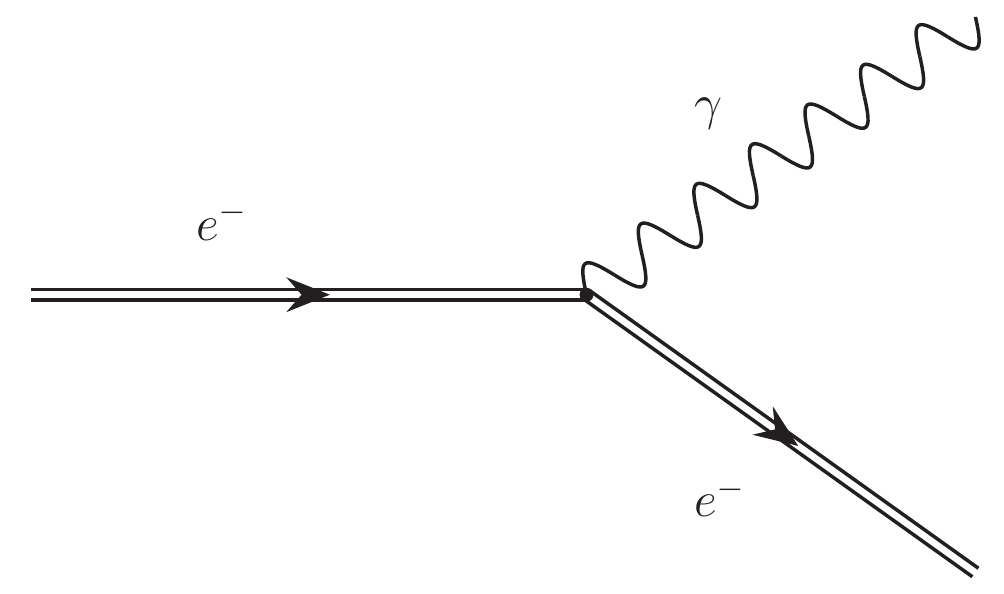}}
\hspace{1cm}
\subfigure[Coherent pair production]{\label{CPP}\includegraphics[width=.35\textwidth]{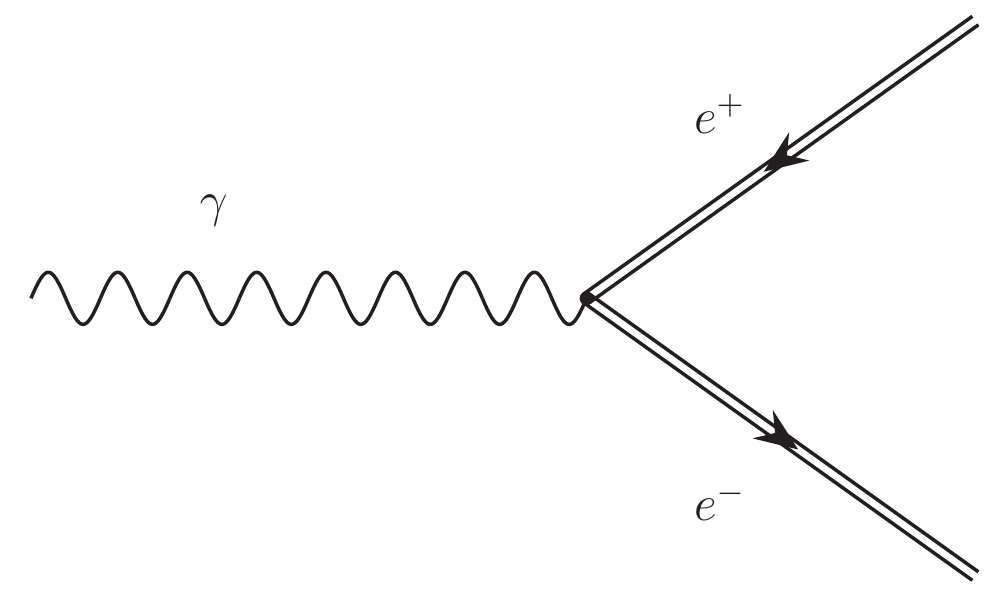}}
\caption{$1^{st}$-order FP processes.}\label{1stOrder}
\end{figure}

The above described Feynman rules are the tool to build every Feynman
diagram that is needed, at each order in perturbation expansion. For
example, beamstrahlung and coherent pair production can be drawn as FP
processes at the $1^{st}$-order in perturbation theory, see
Fig. \eqref{1stOrder}.

\begin{center}
 \begin{figure*}[h]
  \centerline{
    \mbox{\includegraphics[width=5.9in]{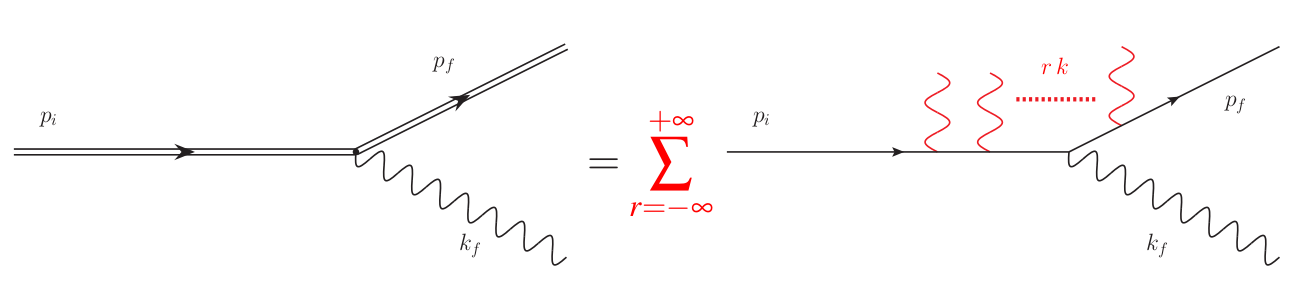}}
      }
  \caption{Na\"{i}ve interpretation of FP beamstrahlung diagram.}
  \label{FP_interpretation}
  \end{figure*}
\end{center}

Writing down the amplitudes of these two processes, one obtains an
expression with a sum coming from the modified QED vertex
\eqref{FP_vert_SUM}. This leads to a na\"{i}ve interpretation
\cite{Nikishov64A} for the new Feynman diagram: it can be seen as sum
over all the Feynman graphs characterized with the emission or absorption
of $r$ photons from the external\footnote{In the case of a constant
crossed field, and hence of a QED vertex with an integral, one can
interpret saying that there has been the absorption/emission either of
one photon with momentum $rk$ or of $r$ photons with momentum $k$.} see
Fig. \eqref{FP_interpretation}.

This interpretation shows that even if the photons of the external field
were not considered at the beginning, they ``appear'' through the
quantum interaction of the fermion with the external field encoded in
the Volkov solution.

Typically, performing probability calculations within the FP, one has to to handle integrals over Airy or Bessel functions coming from the $E_p$ factors,
 that can be simplified using the integral-representation properties of these special functions.


As shortly described in Sec.~\ref{Section_Linacs}, each colliding particle sees
at the IP the superposition of two external fields, but usually only the
field from the oncoming bunch is considered. In order to take fully into
account the effect of the intense fields of \textit{both} colliding
charge bunches at the IP of a collider, new particle wavefunctions are
required. These are obtained by solving the Dirac equation minimally
coupled to the fields of two, non-collinear, constant crossed
fields. Such new solutions will lead to different particle process
transition probabilities. Technically, calculations using the new
solutions may prove simpler than those in one constant crossed field
\cite{Hartin13}.

For an expression of beamstrahlung process through the Furry picture in
$N$ \textit{collinear} constant crossed fields, see
\cite{Hartin:2011vr}.  A new EM solver/generic event generator software,
\texttt{IPStrong}, is being developed to model the strong field
processes at the IP at linear colliders with $\chi$ and the FP-cross
sections as inputs \cite{Hartin13}.

\section{Alternative: generalized quasi-classical operator method (QOM)}\label{Section_QOM}

The conceptually simple FP recipe can lead to relatively complex
analytical calculations already at the first order, due to the --already
mentioned--
multidimensional integrations over special functions and
polynomials. Only a few tree-level two-vertices processes
\cite{HartinThesis, Muller} and one-loop 2-points amplitudes
\cite{Ritus1970} are known, some of them are computed exploiting the
optical theorem, see also, \cite{Kuznetsov:2004tb} and
\cite{Kurilin:1999qc}.

Asymptotic approximations of
results in the FP involving highly energetic particles 
(`ultrarelativistic', i.e.\ Lorentz factor $\gamma_L\gg1$) that are present,
for instance, in the IP region of a linear
collider,  are equivalent to calculations
in a fully quasi-classical approximation \cite{RitusNauka}.

Therefore, Ba\u{\i}er and Katkov invented an effective method
for the calculation of such processes in an external field
using the a quasi-classical approach from the beginning.
Their alternative procedure is well-known as
quasi-classical operator method (QOM) \cite{Baier1968}, for a review
\cite{LandauQED} and \cite{Baier:1998vh}.  The QOM is 
particularly powerful
when considering ultrarelativistic initial state particles and its
results are implemented in \texttt{CAIN} and \texttt{GuineaPig}
\cite{Schulte} to estimate beamstrahlung and coherent pair production at
linear colliders.

The QOM has been originally applied to the case of radiation from a
charged particle in an external magnetic field $\mathbf{B}$ (synchrotron
radiation) \cite{Baier1968}. The field is considered stationary and
$B=|\mathbf{B}|\ll B_{cr}$. From the classical theory the Larmor
frequency $\omega_0$ and the peak frequency of the quasi-continuous
spectrum $\omega$ are such that:
\begin{equation}\label{Larmor_Peak_frequencies}
 \omega_0\approx\frac{|e|\,B}{\epsilon},\,\,\,\,\,\,\,\,\,\,\,\,\,\,\,\,\,\,\,\,\,\,\,\,\,\,\,\,\,\,\,\,\,\,\,\,
\omega\sim\omega_0\left(\frac{\epsilon}{m}\right)^3
\end{equation}
where $\epsilon$ is the energy of the electron.

In this process one can identify two quantum effects: the quantum
propagation of the electron and the quantum recoil on the electron due
to the photon emission.  The relevance of a quantum effect is encoded by
the commutation relations between the operators and by the corresponding
dynamical variables in the uncertainty relations.

Exploiting the non-commutativity between the velocity components of the
  electron in $\mathbf{B}$, one obtains the corresponding
uncertainty
  relations
\begin{equation}
\label{Motion_uncertanties} \Delta
  v_i\,\Delta v_k\sim\frac{e\hbar
  B}{\epsilon^2}=\frac{B}{B_{cr}\gamma^2}=\frac{\hbar\omega_0}{\epsilon},
  \end{equation}
where $\hbar\omega_0$ is the unit energy interval between the possible
electron levels in motion in the field $\mathbf{B}$. Relations
\eqref{Larmor_Peak_frequencies} and \eqref{Motion_uncertanties} show
that the motion is increasingly
classical for increasing energy $\epsilon$.

The non-commutativity between the electron and the emitted photon
dynamical variables is of order
\begin{equation}\label{ElectronPhoton_uncertainties} \frac{\hbar
\omega}{\epsilon}.
                                                                                                  \end{equation}
Considering however, $\chi\sim\frac{\hbar \omega}{\epsilon}\gtrsim1$, it is
obvious that the
classical
theory cannot be applied for the quantum recoil
of the emitted photon energy $\hbar \omega$
\cite{Baier1968}.

Due to \eqref{Motion_uncertanties} and
\eqref{ElectronPhoton_uncertainties}, the key idea of QOM is to consider
the motion of the electron as being classical right from the beginning, 
whereas the quantum recoil from photon emission
is not neglected in the calculation of the amplitude. 
Therefore this  method is called \textit{quasi-classical}.

In order to implement the quasi-classical approach, the amplitudes are
written in terms of \textit{operators} instead of the corresponding
dynamical variables. In particular, the electron solutions of equation
of motion in an external field are not written in terms of Volkov
solutions, but in the symbolic operator form \cite{LandauQED} (operators are
denoted by $\,\hat{\,}\,$):
\begin{equation}\label{Operator_solution}
 \psi(x)=\frac{1}{\sqrt{2\hat{H}}}\,\,u(\hat{P})\,\,e^{-\frac{i}{\hbar}\hat{H}t}\phi(\mathbf{x})
\end{equation}
where $\phi(x)$ is the classical 
wavefunction of a spinless particle and \begin{equation}
                                           u(\hat{P})=\frac{1}{\sqrt{\hat{H}+m}}\begin{pmatrix}
                        \hat{H}+m\\
\mathbf{\sigma}\cdot \hat{\mathbf{P}}
                       \end{pmatrix}
                                          \end{equation}
is the usual Dirac spinor $u(p)$ incorporating the kinematic momentum
operator $\hat{\mathbf{P}}=\hat{\mathbf{p}}-e\mathbf{A}$ of the electron
within the vector potential $\mathbf{A}$ as well as the energy operator
operator $\hat{H}=\sqrt{\hat{\mathbf{P}}^2+m^2}$.  Note, $\hat{\mathbf{P}}$
has not to be misinterpreted as the canonical (or generalized) momentum
$\hat{\mathbf{p}}$ that appears also in the Volkov solutions. In
\eqref{Operator_solution} the information about the external field is
implicitly encoded using $\hat{\mathbf{P}}$. The arbitrary two-component
spinor $w$ satisfies $w^{\ast} w=1$.

The transition matrix element has the form:
\begin{equation}
 U_{fi}=\frac{e\sqrt{2\pi}}{\sqrt{\hbar\omega}}\,\,\int dt\,\,\,\langle f| \,\,\frac{u_f^{\dagger}(\hat{P})}{\sqrt{2\hat{H}}}(\alpha_i\cdot \mathbf{\epsilon^{\ast}_i})\frac{u_i(\hat{P})}{\sqrt{2\hat{H}}}\,\,|i\rangle \,\,\,e^{i\omega t}
\end{equation}
where the bra $\langle f|$ and the ket $|i\rangle$ are solutions of the Klein-Gordon equation in the given field, $\alpha_i=\gamma^0\gamma_i$ ($i=1,2,3$), $\gamma_{\mu}$ are the Dirac matrices and $\epsilon_{\mu}$ the photon polarization vector.

In the probability calculations the operators are kept up  to a later
stage where further commutation relations may appear. According to the
previous prescriptions, the commutation relations of velocity components
of the electron moving in an external field are neglected, while 
the
commutation relations between electron and photon operators 
are left
untouched.

Eventually, one obtains an expression composed by a product of commuting
operators, so that that they can be
substituted by the corresponding classical values (c-numbers).  The
simplification comes in that the relatively complicated expressions
derived from the exact solutions in an external wave do not appear explicitly
while the simple expression for the classical trajectory of the electron
in the field is inserted.

The results by Ba\u{\i}er and Katkov on synchrotron radiation, coherent
pair production and annihilation of a pair into a photon
\cite{Baier1968} are consistent with the ones obtained by Nikishov and
Ritus \cite{Nikishov64A} and previously by Klepikov
\cite{Klepikov:1954}. In the case of the beamstrahlung it has been shown
that the transition probabilities obtained using the FP and the QOM are
asymptotically identical in the ultra-relativistic limit
\cite{Hartin:2009dw}. 
The QOM has been successfully applied to processes
happening in media, in crystals with strong inter-lattice fields and
also in super strong fields (common in astrophysics)
\cite{Baier:1998vh}, \cite{Baier:2009zz}.

The QOM method could in principle be generalized to processes other than
only the photon radiation and cross-symmetric ones. Our present idea is to
understand whether it is possible to apply this kind of approach also to
higher order processes, the main object of studies of future linear colliders, as
it is possible with the Furry picture, and understand whether it is
effective and practical. The process $e^-e^+\rightarrow\mu^-\mu^+$ is
currently under study in the quasi-classical approach. This process has
already been studied in Born approximation \cite{Nedoreshta} and in the
context of laser-driven reaction from positronium using Volkov solutions
\cite{Muller}.

Another promising operator approach, based on the quasi-classical Green's
function of the Dirac equation in an arbitrary is being developed lately
in the context of laser and atomic physics \cite{Milstein1,
DiPiazza110412}.

\section{Conclusions and outlook}\label{Conclusions}

The planned linear collider will produce physics processes in the
environment of very intense electromagnetic fields possibly exceeding the
critical field introduced by Schwinger in  the rest frame of the
colliding electrons and positrons. 

The unstable vacuum present at the
interaction points might lead to a regime of nonlinear Quantum
Electrodynamics, affecting the processes in the IP area.
Such conditions therefore motivate to calculate all  probabilities of
the physics processes under fully consideration of the external
electromagnetic fields affecting the vacuum.

At previous lepton colliders, the much weaker external electromagnetic
fields at the IPs did not needed to be considered apart for background
processes: the first order background processes as beamstrahlung and
coherent pair production, the second order incoherent pair production as
well. At future linear colliders the external fields would be orders of
magnitude higher so an estimate of the effects on all the processes is
requested. As we have shown, indeed, the $\chi$ parameter, that encodes the dependence
 of the probabilities on the intensity of the external field at the IP, is up to 3 orders of magnitude higher at ILC and CLIC than at LEP.
In particular at CLIC-3TeV, we would have $\chi_{av}\sim3.34$, describing a critical regime.

A formally exact method to consider processes in a \textit{classical}
electromagnetic environment is given by the Furry picture of quantum
states. The interaction with the external field is taken into account
not perturbatively and separately from the usual gauge interactions
since fully incorporated in modified equations of motion, the solutions of
which are utilized in the usual S-matrix formalism.

The quasi-classical operator method (QOM) offers an alternative to FP in
the case of ultrarelativistic initial states; a generalization of this
method to two vertex processes is now under study.



\begin{thebibliography}{99}

\bibitem{ATLAS-CONF-2013-014} The ATLAS collaboration, \textit{Mass and Overall Rate Combination of Higgs analyses}, ATLAS-CONF-2013-014.

\bibitem{CMS-PAS-HIG-13-002} The CMS collaboration, \textit{Properties of the Higgs-like boson in the decay $H \to ZZ \to 4l$ in pp collisions at $ \sqrt s =$7 and 8 TeV}, CMS-PAS-HIG-13-002.

\bibitem{ILC-tdr} ILC Global Design Report and Worldwide Study, \textit{ILC Technical Design Report}, to appear soon.

\bibitem{clic-cdr} \textit{Physics and Detectors at CLIC: CLIC Conceptual Design Report}, edited by L.~Linssen, A.~Miyamoto, M.~Stanitzki, H.~Weerts, CERN-2012-003.

\bibitem{Klein} O.~Klein, \emph{Die Reflexion von Elektronen an einem Potentialsprung nach der relativistischen Dynamik von Dirac},
\emph{Z. Phys.} {\bf 53} (1929) 157.

\bibitem{Sauter} F.~Sauter, \emph{\"{U}ber das Verhalten eines Elektrons im homogenen elektrischen Feld nach der relativistischen Theorie Diracs}, \emph{Z. Phys.} {\bf 69} (1931) 742.

\bibitem{Schwinger}   J.~Schwinger, \emph{On Gauge Invariance and Vacuum Polarization}, \emph{Phys. Rev.} {\bf 82} (1951) 664.

\bibitem{FradkinBook} E.S.~Fradkin, D. M. ~Gitman and Sh.M.~Shvartsman, \textit{Quantum electrodynamics with unstable vacuum}, Springer, Berlin 1991.

\bibitem{Kuznetsov:2004tb} A.~Kuznetsov and N.~Mikheev,\textit{ Electroweak processes in external electromagnetic fields}, \textit{Springer Tracts Mod.Phys.} \textbf{197} (2004).




\bibitem{Greiner} W.~Greiner, B.~M\"{u}ller and J.~Rafelski, \textit{Quantum electrodynamics of strong fields}, Springer, Berlin 1985.

\bibitem{XFEL} \textit{The European X-Ray Free-Electron Laser: Technical Design Report}, DESY 2006-097.

\bibitem{Bamber} C.~Bamber \textit{et al.}, \emph{Studies of nonlinear QED in collisions of 46.6 GeV electrons with intense laser pulses}, \emph{Phys.Rev. D} {\bf60} (1999) 092004. 


\bibitem{Heinzl} T.~Heinzl and A.W.~Ilderton, \emph{Extreme field physics and QED}, {\tt hep-ph/0809.3348}.

\bibitem{DiPiazza12} A.~Di Piazza, C.~M\"{u}ller, K.Z.~Hatsagortsyan and C.H.~Keitel, \emph{Extremely high-intensity laser interactions with fundamental quantum systems}, \emph{Rev.Mod.Phys.} {\bf 84} (2012) 1177.

\bibitem{Furry} W.H.~Furry, \emph{On Bound States and Scattering in Positron Theory}, \emph{Phys.Rev.} \textbf{81} (1951) 115.

\bibitem{YokoyaChen} K.~Yokoya and P.~Chen, \textit{Beam-beam phenomena in linear colliders}, \textit{Lect.Notes Phys.} \textbf{400} (1992) 415.

\bibitem{Baier1968} V.N.~Ba\u{\i}er and V.M.~Katkov, \textit{Processes Involved in the Motion of High Energy Particles in a Magnetic Field}, \textit{Sov.Phys. JETP} \textbf{26} (1968) 854.

\bibitem{EPA} C.F.~Weizsacker, \textit{Ausstrahlung bei St\"{o}\ss{}en sehr schneller Elektronen}, \textit{Z. Phys.}, \textbf{88} (1934) 612; E.J.~Williams, \textit{Nature of the High Energy Particles of Penetrating Radiation and Status of Ionization and Radiation Formulae}, \textit{Phys. Rev.} \textbf{45} (1934) 729.

\bibitem{Muller} C.~M\"uller, K.Z.~Hatsagortsyan, and C.H.~Keitel, \textit{Muon pair creation from positronium in a circularly polarized laser field}, \textit{Phys. Rev. D} \textbf{74} (2006) 074017, [\texttt{physics/0602106}] ; C.~M\"uller, K.Z.~Hatsagortsyan, and C.H.~Keitel, \textit{Muon pair creation from positronium in a linearly polarized laser field}, \textit{Phys. Rev. A} \textbf{78} (2008) 033408, [\texttt{physics/0804.4813}]; C.~M\"uller, K.Z.~Hatsagortsyan, and C.H.~Keitel, \textit{Particle physics with a laser-driven positronium atom}, \textit{Phys. Lett. B} \textbf{659} (2008) 209, [\texttt{hep-ph/0705.0917}].

\bibitem{Kurilin:1999qc} A.V.~Kurilin, \textit{Particle physics in intense electromagnetic fields}, \textit{Nuovo Cim. A} \textbf{112} (1999) 977.

\bibitem{Nikishov64B} A.I.~Nikishov and V.I.~Ritus, \emph{Quantum Processes in the Field of a Plane Electromagnetic Wave and in a Constant Field 2}, \textit{Sov.Phys. JETP} \textbf{19} (1964) 1191.

\bibitem{RitusNauka} V.I.~Ritus, \emph{Quantum effects of the interaction of elementary particles with an intense electromagnetic field}, \emph{J. Sov. Las. Res.} {\bf 6} (1985) 497.


\bibitem{King} B.~King, N.~Elkina and H.~Ruhl, \textit{Photon polarisation in electron-seeded pair-creation cascades}, \texttt{
hep-ph/1301.7001}; B.~King and H.~Ruhl, \textit{Trident pair production in a constant crossed field}, \texttt{hep-ph/1303.1356}.



\bibitem{CAIN} K.~Yokoya, \textit{User's Manual of CAIN version 2.35}, \textit{KEK Pub} 4/96 (2003).

\bibitem{LandauQED} V.B.~Berestetskii, E.M.~Lifshitz, and L.P.~Pitaevskii, \textit{Quantum Electrodynamics - 2nd Edition}, Pergamon Press, Oxford 1982.

\bibitem{MoortgatPick:2009zz}
  G.~Moortgat-Pick,
  J.\ Phys.\ Conf.\ Ser.\  {\bf 198} (2009) 012002.


\bibitem{Volkov} D.M.~Wolkow, \emph{\"{U}ber eine Klasse von L\"{o}sungen der Diracshen Gleichung},  \textit{Z. Phys.} \textbf{94} (1935) 250.

\bibitem{Ritus1970} V.I.~Ritus, \textit{Radiative corrections in quantum electrodynamics with intense field and their analytical properties }, \textit{Ann. Phys.} \textbf{69} (1970) 555.

\bibitem{Boca} M.~Boca and V.~Florescu, \textit{The completeness of Volkov spinors}, \textit{Rom. Jour. Phys.} \textbf{55} (2010) 511.

\bibitem{Nikishov64A} A.I.~Nikishov and V.I.~Ritus, \emph{Quantum Processes in the Field of a Plane Electromagnetic Wave and in a Constant Field 1}, \textit{Sov.Phys. JETP} \textbf{19} (1964) 529.



\bibitem{Hartin13} A.~Hartin, in preparation (2013).

\bibitem{Hartin:2011vr} A.~Hartin and G.~Moortgat-Pick, \textit{High Intensity Compton Scattering in a strong plane wave field of general form}, \textit{Eur.Phys.J. C} \textbf{71} (2011) 1729.


\bibitem{HartinThesis} A.~Hartin, \textit{Second order processes in an intense electromagnetic field}, PhD thesis,
University of London, 2006.




\bibitem{Baier:1998vh} V.N.~Ba\u{\i}er, V.M.~Katkov and V.M.~Strakhovenko, \textit{Electromagnetic processes at high energies in oriented single crystals}, World Scientific Publishing Company, Singapore 1998.


\bibitem{Schulte}   D.~Schulte, \textit{Study of Electromagnetic and Hadronic Background in the Interaction Region of the TESLA Collider}, PhD Thesis, \texttt{TESLA 1997-08}.




\bibitem{Klepikov:1954} N.P.~Klepikov, \textit{Emission of photons and electron-positron pairs in a magnetic field}, \textit{Zh. Eksp. Teor. Fiz.} \textbf{26} (1954) 19.
   

\bibitem{Hartin:2009dw} A.F.~Hartin, \textit{On the equivalence of semi-classical methods for QED at the Interaction Point of future linear colliders}, \textit{J. Phys. Conf. Ser.} \textbf{198} (2009) 012004.

\bibitem{Baier:2009zz} V.N.~Baier and V.M.~Katkov, \textit{Recent development of quasiclassical operator method}, \textit{J.Phys.Conf.Ser.} \textbf{198} (2009) 012003.



\bibitem{Nedoreshta} V.N.~Nedoreshta, A.I.~Voroshilo and S.P.~Roshchupkin,  \textit{Nonresonant scattering of an electron by a muon in the field of plane electromagnetic wave}, \textit{Laser Phys. Lett.} \textbf{4} (2007) 872; V.N.~Nedoreshta, A.I.~Voroshilo and S.P.~Roshchupkin, \textit{Resonant scattering of an electron by a muon in the field of light wave}, \textit{Eur. Phys. J. D} \textbf{48} (2008) 451.


\bibitem{Milstein1} A.I.~Mil'shtein and V.M.~Strakhovenko, \textit{Quasiclassical approach to the high-energy Delbr\"{u}ck scattering}, \textit{Phys. Lett. A}  \textbf{95} (1983) 135.

\bibitem{DiPiazza110412} A.~Di Piazza and A.I.~Milstein, \textit{Quasiclassical approach to high-energy QED processes in strong laser and atomic fields}, \textit{Phys.Lett. B} \textbf{717} (2012) 224 \texttt{[physics.atom-ph/1204.2502]}; A.~Di Piazza, \textit{On refractive processes in strong laser field quantum electrodynamics}, \texttt{hep-ph/1303.5353}.








\end{thebibliography}
\end{document}